\documentclass[usenatbib]{mn2e}
\usepackage{lipsum}
\usepackage[usenames,dvipsnames,svgnames,table]{xcolor}
\usepackage{graphicx}
\usepackage{color}

\newcommand{\mytilde}{\raise.17ex\hbox{$\scriptstyle\mathtt{\sim}$}}

\title[]
  {{\em Swift} J1112.2-8238: A Candidate Relativistic Tidal Disruption Flare}
\author[G.Brown et al.]
  {G.C.~Brown$^1$\thanks{E-mail:- g.c.brown@warwick.ac.uk}, A.J.~Levan$^1$, E.R.~Stanway$^1$, N.R. Tanvir$^2$,  S.B. Cenko$^3$,  E.Berger$^4$, \and R. Chornock$^5$, A. Cucchiaria$^3$  
  \newauthor \\
  $^1$Department of Physics, University of Warwick, Gibbet Hill Road, Coventry CV4 7AL, UK\\
  $^2$Department of Physics and Astronomy, University of Leicester, University Road, Leicester, LE1 7RH, UK \\
  $^3$Astrophysics Science Division, NASA Goddard Space Flight Center, Mail Code 661, Greenbelt, Maryland 20771, USA \\
  $^4$Harvard-Smithsonian Center for Astrophysics, 60 Garden Street, Cambridge, MA 02138, USA \\
  $^5$Department of Physics and Astronomy, Ohio University, 251B Clippinger Lab, Athens, OH 45701, USA\\
    }
\date{May 2015}

\pagerange{\pageref{firstpage}--\pageref{lastpage}} \pubyear{2002}

\def\LaTeX{L\kern-.36em\raise.3ex\hbox{a}\kern-.15em
    T\kern-.1667em\lower.7ex\hbox{E}\kern-.125emX}

\begin{document}

\label{firstpage}

\maketitle
\begin{abstract}

We present observations of {\em Swift} J1112.2-8238, and identify it as a candidate relativistic tidal disruption flare (rTDF). The outburst was first detected by {\em Swift}/BAT in June 2011 as an unknown, long-lived (order of days) $\gamma$-ray transient source. We show that its position is consistent with the nucleus of a faint galaxy for which we establish a likely redshift of $z=0.89$ based on a single emission line that we interpret as the blended [O{\sc ii}]$\lambda3727$ doublet. At this redshift, the peak X/$\gamma$-ray luminosity exceeded $10^{47}$ ergs s$^{-1}$, while a spatially coincident optical transient source had $i^{\prime} \mytilde 22$ (M$_g \mytilde -21.4$ at $z=0.89$) during early observations, \mytilde 20 days after the {\em Swift} trigger. These properties place {\em Swift} J1112.2-8238 in a very similar region of parameter space to the two previously identified members of this class, {\em Swift} J1644+57 and {\em Swift} J2058+0516. As with those events the high-energy emission shows evidence for variability over the first few days, while late time observations, almost 3 years post-outburst, demonstrate that it has now switched off. {\em Swift} J1112.2-8238 brings the total number of such events observed by {\em Swift} to three, interestingly all detected by {\em Swift} over a $\mytilde$3 month period ($<3\%$ of its total lifetime as of March 2015). While this suggests the possibility that further examples may be uncovered by detailed searches of the BAT archives, the lack of any prime candidates in the years since 2011 means these events are undoubtedly rare.

\end{abstract}


\section{Introduction}\label{sec:int}

In recent years it has been assumed that the majority, if not all, large galaxies house at their cores a supermassive black hole (SMBH) ranging from many hundreds of thousands to billions of times the mass of our Sun  \citep[see the recent review by][]{Graham2015}. These objects strongly influence, or are strongly influenced by, the properties of their hosts, as evidenced by certain galaxy-wide properties scaling with the masses of these SMBHs, such as in the M-$\sigma$ relation \citep{Ferrarese2000,Gebhardt2000}. However, the cause of this link is not well understood. In order to better understand the SMBH demographic \citep[particularly at the low mass end where samples are of very limited size, see e.g.][]{Reines2013}, and their co-evolution with their hosts, further examples must be studied in dwarf and distant galaxies. 

However, in the case of a distant/dwarf galaxy lacking an active galactic nucleus (AGN), even confirming the existence of an SMBH can be difficult. Obtaining spatially resolved velocity dispersion measurements across the the galaxy, looking for the gravitational influence of a massive central body, becomes impossible with current instrumentation when the angular size of the galaxy becomes too small. The detection and correct identification of a tidal disruption flare (TDF), however, unequivocally shows the existence of an SMBH within the flare's host, irrespective of the host's angular size or apparent magnitude.

A TDF is the luminous burst produced by the capture, disruption and
subsequent accretion of a star onto a SMBH. They are typically
characterised by a short-lived (months to years) transient with a high
temperature ($>10^4$ K) thermal spectral energy distribution
(SED). They occur whenever a star passes within its tidal radius of
the central SMBH ($r_{\mathrm{t}} \sim R_* (M_{\mathrm{BH}} /
M_*)^{1/3}$) while remaining outside of the Schwarzschild radius
($R_{\mathrm{S}}$) \citep{Rees1988}, since crossing the latter would
lead to the star being swallowed whole and thus produce no visible
flare. Since $r_{\mathrm{t}} \propto M_{\mathrm{BH}}^{1/3}$ and
$R_{\mathrm{S}} \propto M_{\mathrm{BH}}$, for a given radius (and
mass) of star, there exists a maximum black hole mass for which the
disruption will occur outside $R_S$ \citep[although in practice the
  spin of the black hole is also important,
  e.g. ][]{Kesden2012a}. Hence white dwarfs
  will only be disrupted by intermediate mass black holes ($\sim 10^5$
  M$_{\odot}$), while main sequence stars can produce
  flares with black holes up to $\sim 10^8$ M$_{\odot}$ and giants may
  be disrupted even around the most massive known black holes, though
  the likely accretion rates and timescales are much longer than for
  more compact systems \citep{MacLeod2013}. Ultimately, observations
  of confirmed tidal flares may offer a new approach
    to measurements of black hole masses
  \citep{Lodato2011,Gezari2012} and spins \citep{Kesden2012b}.

A number of TDF candidates have been found in the UV \citep[e.g. ][]{Gezari2008, Gezari2009, Gezari2012}, soft X-rays \citep[e.g. ][]{Brandt1995,Grupe1995,Bade1996,Cappelluti2009} and the optical \citep[e.g. ][]{vanVelzen2011,Cenko2012b,Arcavi2014}. However, most of these were detected at low redshift, often less than $z=0.1$ due to the necessity of multi-epoch photometry and astrometry, and the relatively shallow surveys from which they were selected. Probing these events to much greater distances, thus providing a way to characterise the SMBH mass distribution as a function of redshift, would require much more sensitive, high cadence surveys \citep[such as the LSST, ][]{Izevic2014} or possibly chance gravitational lensing events \citep[e.g. the z=3.3 candidate, ][]{Stern2004}. However, a new sub-class of these events, potentially observable out to much larger distances with current observing platforms, has provided us with a new way to observe these transients. 

The first such event {\em Swift} J164449.3+573451 (henceforth {\em Swift} J1644+57), detected in 2011 March, exhibited extremely unusual high energy behaviour. Detected initially as a gamma-ray burst (GRB) trigger with a long duration  \citep[$\mytilde$1000s, ][]{Cummings2011}, the event remained bright and variable for several days, retriggering {\em Swift}/BAT on a further three occasions over the course of 48 hours \citep[][]{Cummings2011b} making it clear this was not a standard GRB, short or long. X-ray monitoring with {\em Swift}/XRT showed a luminous flaring source that settled into a several-day long plateau before following an approximately power law decay, all with considerable short-term variability superimposed upon it. The source was discovered to lie at a cosmological distance, coincident with the centre ($<150$ pc to $1\sigma$) of a faint star-forming host at a spectroscopically confirmed redshift of $z=0.353$ \citep[][]{Levan2011}, implying the isotropic X-ray luminosity of the event was $L_{\mathrm{X}} = 10^{47}-10^{48} \mathrm{erg\,s^{-1}}$ even at late times. In contrast, the coincident optical/infrared transient peaked at a more modest $L_{\mathrm{opt/IR}} = 10^{42}-10^{43} \mathrm{erg\,s^{-1}}$ even after correction for moderate internal extinction \citep[][]{Bloom2011}.

Radio observations of {\em Swift} J1644+57 with the EVLA detected a rising unresolved source with an equipartition radius that implied a moderately relativistic expansion with Lorentz factor $\Gamma \mytilde 2$ and a formation epoch that coincided with the initial $\gamma$-ray detection \citep[][]{Zauderer2011}. The energetics measured by \citet[][]{Zauderer2011}, $\mytilde 3 \times 10^{50}$ erg at 22 days post burst, also corresponded to the Eddington luminosity for accretion onto a $10^6 \mathrm{M_{\odot}}$ black hole. \citet[][]{Miller2011} used a purely observational relation between X-ray luminosity, radio luminosity and the mass of black holes (ranging from high mass Seyferts to stellar mass black holes) to estimate the mass of the SMBH that produced {\em Swift} J1644+57. The resulting weak constraint of $\log(M_{BH}/M_{\odot}) = 5.5 \pm 1.1$, was consistent with the black hole mass of $2\times10^{6} - 10^{7}$ estimated by \citet[][]{Levan2011} via the spheroid mass-black hole mass scaling relation of \citet[][]{Bennert2011}.

These unique broadband properties marked {\em Swift} J1644+57 as a new class of transient. They suggested the detection of a flare situated in the nuclear region of a dwarf galaxy with energetics and short-term variability consistent with an accretion event onto the central SMBH. But the lack of any previous activity in $\gamma$-rays during the lifetime of Swift \citep[][]{Krimm2011GCN}, the spectroscopic classification of the galaxy as star-forming \citep[][]{Levan2011} and the radio formation epoch \citep[][]{Zauderer2011} all indicated the accretion event was new and not part of any ongoing nuclear activity. Thus, the favoured explanation was taken to be that of the tidal disruption of a solar-type star that had also launched a moderately relativistic jet. However, this relativistic tidal disruption flare (rTDF) interpretation was not unchallenged, and other mechanisms, perhaps involving
the tidal capture of a white dwarf \citep{Krolik2011} or massive star core collapse \citep{Quataert2012,Woosley2012}, were postulated.

A second example, {\em Swift} J2058.4+0516 \citep[{\em Swift} J2058+05, ][]{Cenko2012}, was detected in May 2011. While apparently much fainter than {\em Swift} J1644+57, this was largely due to the much greater redshift ($z=1.19$, c.f. $z=0.35$ for {\em Swift} J1644+57) and this was in fact a more luminous event. The bulk properties (peak luminosity, total energy, longevity, steep late-time cut-off) of the event matched well with those of {\em Swift} J1644+57 \citep[][]{Pasham2015} although there were several important differences. The X-ray lightcurve decline was steeper (a power law with index $\mytilde-2.2$, while {\em Swift} J1644+57 was remarkably near the theoretical $-5/3$ \citep[][]{Rees1988,Phinney1989}, although recent numerical simulations suggest that accounting for stellar structure and closeness of approach, -2.2 is expected in half of all disruptions \citep{guillochon2013}) and the X-ray spectrum was somewhat harder (photon index $\mytilde1.6$, c.f. $\mytilde2$). In addition, the observed radio spectrum of {\em Swift} J2058+05 was very flat ($\nu^{0}$) which constrasts strongly with the optically thick spectrum ($\nu^{1.3}$) of {\em Swift} J1644+57 \citep[][]{Pasham2015}. Despite these differences, it was suggested that {\em Swift} J2058+0516 is the second member of the rTDF class, and in this case the observational differences may offer important diagnostics of the disruption process.

Another potentially related class of transient is that of the ultra-long GRBs (ULGRBs, \cite{Levan2014}). These events exhibit $\gamma-$ray emission lasting for thousands of seconds (1-2 orders of magnitude less than the flares described above, but an order of magnitude longer than most
GRBs). While multiple possible paths to their creation have been suggested \citep[e.g.][]{Thone2011,Campana2011,Gendre2013}, it is also
possible that they are related to TDFs with a relativistic component, although in this case their shorter timescales would imply a white dwarf disruption \citep{Levan2014,Krolik2011,MacLeod2014}.

Relativistic TDFs are potentially observable out to much greater distances than their non-relativistic thermal TDF cousins due to their beamed emission, analogous to how GRBs have been shown to be detectable to extreme redshifts \citep[e.g. GRB090423, ][]{Tanvir2009,Salvaterra2009}. \citet[][]{Levan2011} estimated that {\em Swift} J1644+57 would have been observable out to $z>0.6$, while the redshift of {\em Swift} J2058+05 \citep[][]{Cenko2012} ($z=1.185$) shows certain members may be observable at even larger distances. \citet[][]{Zauderer2011} suggest that the radio component would be detectable out to $z \sim 6$, potentially making large scale radio surveys a powerful method for the detection of these events.

Such rTDFs also evolve on very short timescales compared to AGN and so offer a way to study jetted accretion events across their whole lifetimes on human timescales, evidenced by observations of {\em Swift} J1644+57 showing that the jet has now apparently shut-off \citep[][]{Zauderer2013}. This is a virtual impossibility in the vastly longer lived AGN duty cycles which may last in excess of $10^{7}$ years \citep[e.g.][]{Hopkins2009}. In addition, the relativistic jets emitted by these events have been suggested as a possible source of ultra-high energy cosmic rays \citep[e.g.][]{Farrar2009,Bloom2011, Wang2011,Farrar2014}.

Given the potential importance of studying these events, it is concerning that it is unclear whether {\em Swift} J2058+05 would have attracted such detailed follow-up in the absence of {\em Swift} J1644+57, considering to its less immediately impressive nature. In addition, the notable temporal coincidence of the two bursts, being only two months apart in the then $\mytilde7$ years that Swift had been operating, led to the suggestion at the time that further examples within the {\em Swift} archive may have been overlooked \citep[e.g.][]{Levan2011}.

Here we present observations of {\em Swift} J1112.2-8238 (henceforth {\em Swift} J1112-8238) which was detected by the {\em Swift} Burst Alert Telescope in 2011 June and whose nature has to date been uncertain. Our spectroscopic observations establish a cosmological redshift for the transient, and we demonstrate optical variability close to the nucleus of a faint galaxy. We analyse the inferred physical properties, comparing them to the properties of previous {\em Swift} flare classifications including the recently discovered ultra-long GRBs, and to the established relativistic TDF candidates, leading us to suggest that {\em Swift} J1112-8238
is also a candidate rTDF. 

All magnitudes presented in this paper are in the AB magnitude system.  Where
necessary, we use a standard $\Lambda$CDM cosmology with
$H_0=$70\,km\,s$^{-1}$\,Mpc$^{-1}$, $\Omega_M=0.3$ and
$\Omega_\Lambda=0.7$.

\section{Observations}\label{sec:obs}

\subsection{Swift BAT data}

The outburst of {\em Swift} J1112-8238 was originally discovered by the {\em Swift} telescope \citep{Gehrels2004} in a four day integration by the Burst Alert Telescope \citep[BAT, ][]{Barthelmy2005} between 2011 June 16 and 19 \citep[MJD 55728-55731,][]{Krimm2011ATel}. We choose to set the trigger time to 2011 June 16 {\sc ut} 00:01, although in practice a precise trigger time is poorly defined. The count rate in gamma rays across this period was $(2.9 \pm 0.7) \times10^{-3}$\,ph s$^{-1}$\,cm$^{-2}$ with a peak daily average rate of $(1.9 \pm 0.5) \times 10^{-2}$\, ph s$^{-1}$\,cm$^{-2}$ recorded on the 16th \citep[][]{Krimm2011ATel}. This peak, though high, was not in itself sufficient to trigger the automated transient monitor which has a $5\sigma$ burst detection threshold \citep[][]{Krimm2011ATel,Krimm2013}. We utilise the available BAT daily average light curves\footnote{http://swift.gsfc.nasa.gov/results/transients/} extending back as far as the launch of {\em Swift} in 2005 and confirm that there was no pre-trigger or post-burst activity above a $5\sigma$ threshold level at the flare's position over any 4-day window. We also note that \cite{Krimm2013} report no evidence for additional flares from the source. The light curve ($\gamma$-ray, X-ray, optical) is shown in Figure 1.

\subsection{X-ray data}

Initial X-ray data was obtained by the {\em Swift} X-Ray Telescope \citep[XRT, ][]{Burrows2005} in a 3000s target of opportunity observation approximately 10 days after the initial trigger \citep[MJD 55741.7,][]{Krimm2011ATel}. The source was well detected with an observed flux of $1.4^{+0.1}_{-0.1}\times10^{-11}$ ergs s$^{-1}$cm$^{-2}$. An X-ray monitoring programme continued for a further 30 days with all observations obtained in photon counting (PC) mode. We obtained reduced XRT products from the {\em Swift} archive\footnote{http://www.swift.ac.uk/user$\_$objects}, created using the techniques outlined in \cite{Evans2007,Evans2009,Evans2010}. 
The enhanced X-ray position derived from the UVOT boresight correction is RA= 11:11:47.32  DEC=-82:38:44.2 (J2000) with a 90\% error radius of $1.4^{\prime\prime}$. 
The combined spectrum of all available PC-mode data is well fit by an absorbed power-law of photon index $\Gamma_{\mathrm{ph}}=1.33 \pm 0.08$ and N$_{\mathrm{H}}$ (int) = $2.4^{+1.8}_{-1.6}\times10^{21}$ cm$^{-2}$ consistent with the Galactic value of $1.8\times10^{21}$cm$^{-2}$ \citep{Willingale2013}. Although the number of counts is small, there is no evidence for spectral evolution through the observations. 

The light curve over the same period exhibits a gradual decay but with marked variability (a factor of $\mytilde2$ in flux) between individual
snapshots (uninterrupted pointings). We obtained further late time observations in April 2014, with a total XRT exposure
time of 6960.3 s (in PC mode). This observation provides an upper limit on the source flux of $F_{\mathrm{X}} < 4 \times 10^{-14}$ ergs s$^{-1}$ cm$^{-2}$ \citep[99\%, determined
via the Bayesian method of][]{Kraft91}. This is a factor of $\mytilde 250$ fainter than the peak luminosity confirming the source's transient nature.

\begin{figure}
 \includegraphics[width=8.4cm]{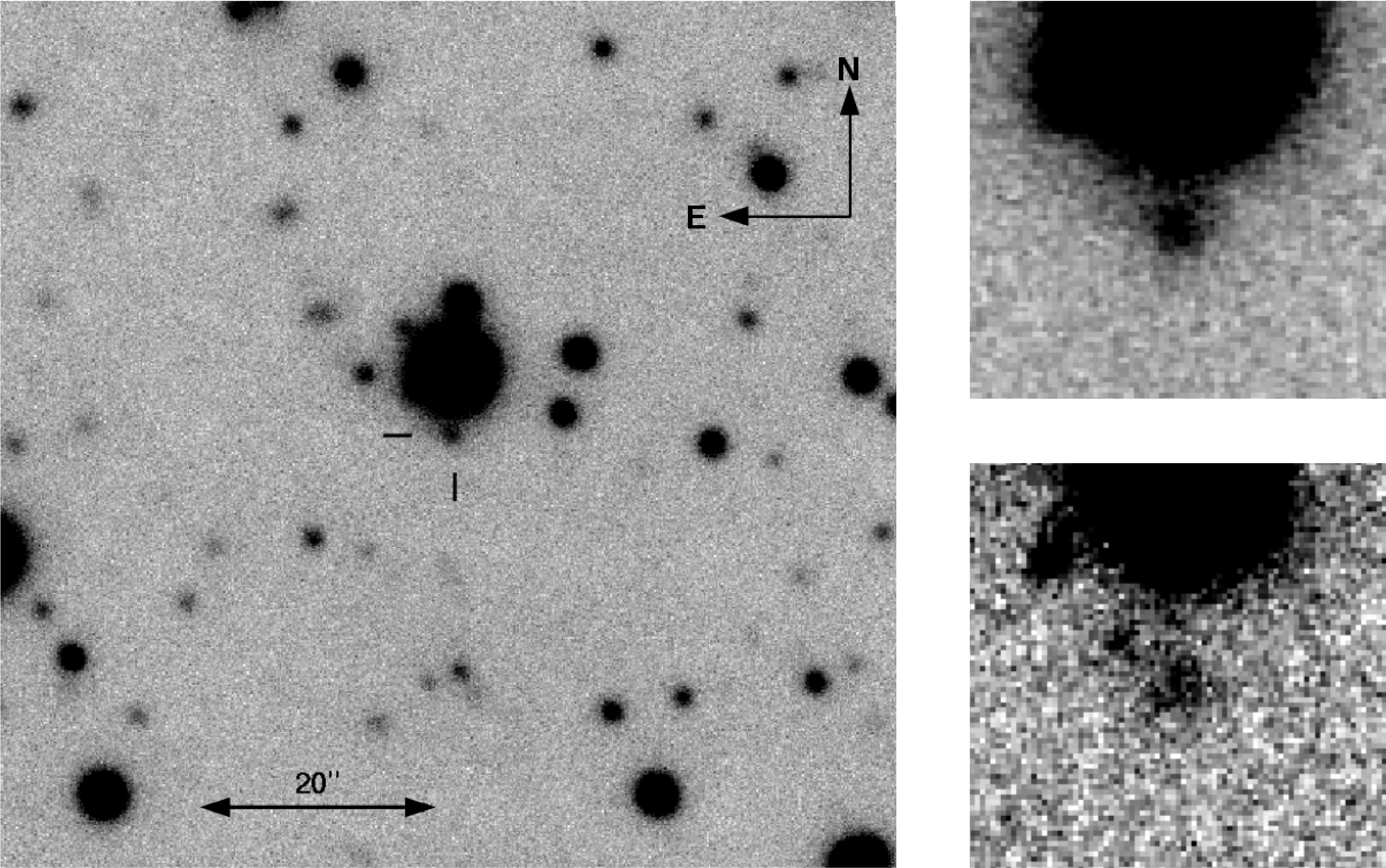}
 \caption{(Left) A GMOS-S $i'$-band finding chart for {\em Swift} J1112-8238. (Right) A comparison between the source at $\mytilde$20 days (Top) and at $\mytilde$1.5 years (Bottom) post trigger, each panel 15$^{\prime\prime}$ across. The extended host's structure is far clearer in the later epoch, due in part to both the decline of the optical transient and the greatly improved seeing. \label{fig:finding}}
\end{figure}

\begin{figure*}
 \includegraphics[width=16.8cm]{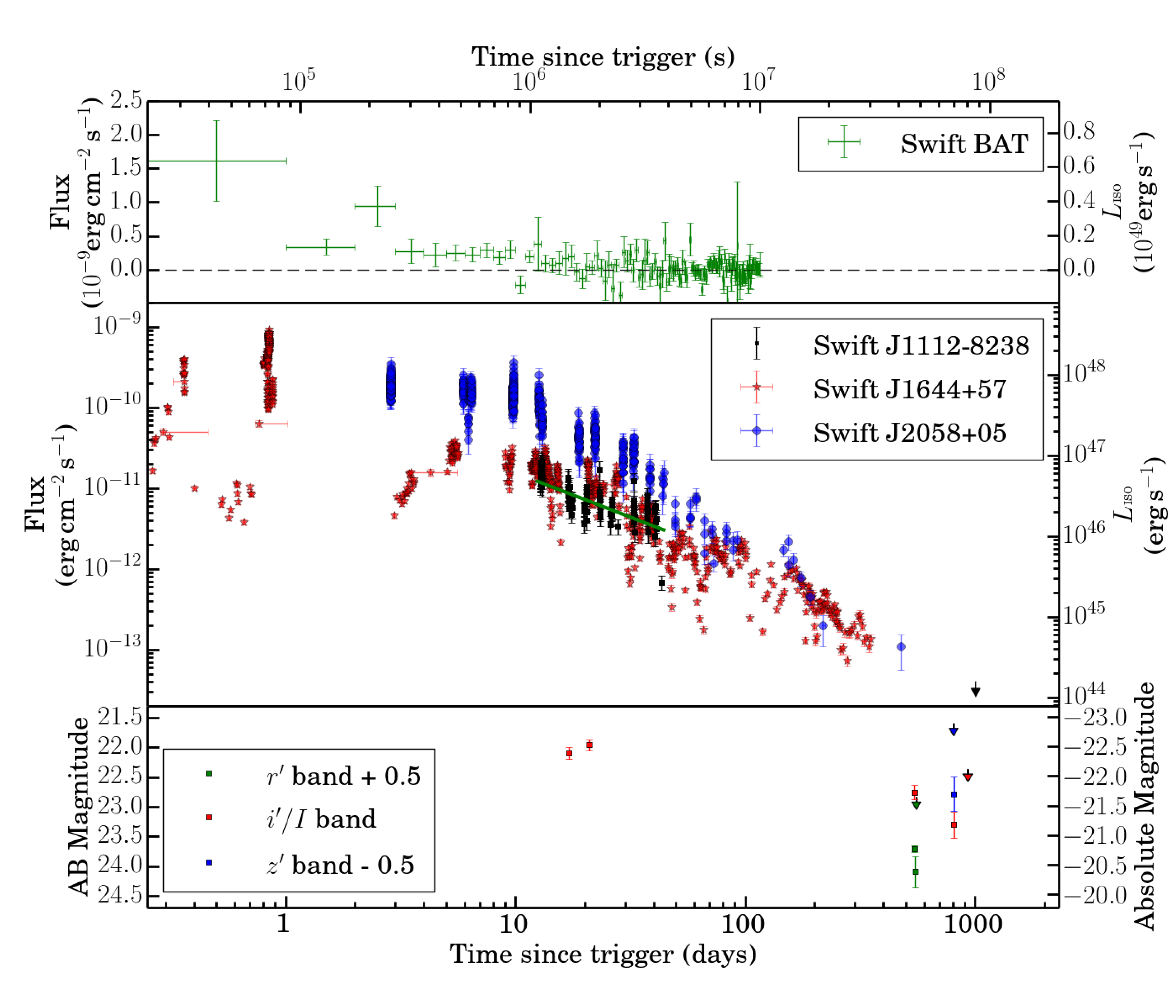}
 \caption{The lightcurves for {\em Swift} J1112-8238 in various wavebands from a few hours to $\mytilde 1000$ days post-trigger. The time axis is displayed in both seconds and days in the observer frame. In all panels, the right hand axis indicates an isotropic equivalent luminosity or equivalent optical absolute magnitude. (Top) The median subtracted {\em Swift}/BAT daily average lightcurve in the 15-150 keV range, cut at $10^{7}$ seconds post trigger for clarity. Note that the vertical scale is linear, and there are no significant detections beyond the first few days. (Middle) The {\em Swift}/XRT lightcurve in the 0.2-10 keV range. The green line indicates a t$^{-1.1}$ fit on the data preceding the sharp decline at $\mytilde30$ days post trigger. In addition, the X-ray luminosities for {\em Swift} J1644+57 and {\em Swift} J2058+05 have been plotted. To allow for a direct comparison between the lightcurves, {\em Swift} J1644+57's and {\em Swift} J2058+05's lightcurves have had a cosmological time dilation correction to place them as though they had occured at the same redshift as {\em Swift} J1112-82. (Bottom) The optical lightcurves from GMOS-S and FORS2 photometry. There is considerable optical variability between the early and late time $i^{\prime}$/$I$ band magnitudes and the late time (\textgreater 1.5 year) magnitudes are assumed to be at host level. \label{fig:lightcurve}  }
\end{figure*}

\subsection{Optical Imaging}

Following a UVOT non-detection \citep[b $> 22.0$ mag, ][]{Krimm2011ATel} made at the beginning of the X-ray monitoring programme, we obtained observations in the $i^{\prime}$ band with the Gemini Multiple Object Spectrograph on Gemini South \citep[GMOS-S,][]{Hook2004} at 2011 July 3 {\sc ut} 00:58, 17 days after initial trigger \citep[][]{Berger2011ATel}. We performed later follow-up with GMOS-S in the $r^{\prime}$ and $i^{\prime}$-bands at 1.5 years post-trigger (starting 2012 December 13 {\sc ut} 06:50), and with the FOcal Reducer and low dispersion Spectrograph 2 (FORS2) on the Very Large Telescope (VLT)  at 2 years post-trigger (starting 2013 August 31 {\sc ut} 23:31) in $I$ and $z^{\prime}$. In addition, as part of the GMOS-S spectroscopic follow-up, a number of short exposure acquisition images were taken in $r^{\prime}$ (starting at 2012 December 16 {\sc ut} 07:30 and 2012 December 23 {\sc ut} 05:16) and in $i^{\prime}$ (starting at 2014 January 3 {\sc ut} 07:01, $\mytilde 3$ years post trigger). All of this imaging was reduced using standard {\sc iraf} and {\sc esorex} data reduction techniques. We note that the presence of a nearby bright star (R = 15.8 mag at an angular distance of $\mytilde5^{\prime\prime}$, Figure \ref{fig:finding}) complicated the analysis of this source. We remove the majority of the flux from this contaminating star by subtracting a model stellar PSF, constructed as a median-averaged radial light profile for the star in question. We also considered models for the PSF from stars elsewhere in the image, subtracting a rotated copy of the contaminating star, or modelling the star directly via Moffat or multiple Gaussian fits. All these methods were hampered by the existence of other objects close to the star, or left clear residuals in the data.  

Photometric calibrations for the $i^{\prime}$/$I$ band was completed through comparison with observations of photometric standards analysed via the {\sc esorex} FORS2 pipeline, the expected systematic offset between the GMOS $i^{\prime}$ and FORS2 $I$ filters having been deemed negligible in this low signal to noise regime. The non-standard filter $z^{\prime}$ was instead calibrated through comparison with the FORS2 standard star, Feige 110, which was observed within a few nights of our observations. Finally the $r^{\prime}$ band was calibrated with reference to the Gemini standard zeropoints \footnote{http://www.gemini.edu/sciops/instruments/gmos/calibration/ photometric-stds}.

The resultant photometry is detailed in Table \ref{tab:Photom} and plotted in Figure \ref{fig:lightcurve}. The early time observations showed a point-like source while later observations ($>$1\,yr) reveal emission with a flux a factor $\mytilde 2$ lower than recorded at early times. Modelled photometry of the late time emission with a S\'{e}rsic profile using {\sc galfit} \citep[][]{Peng2002,Peng2010} was consistent with the aperture photometry detailed above, while PSF-matched point-source photometry (completed by scaling a PSF built from the image) yields results a magnitude dimmer, indicating the late time source is extended.

The photometry has been corrected for Galactic extinction, with E(B-V) $= 0.253\pm0.009$, based on values derived from \citet[][]{Schlafly2011} and accessed via the NASA/IPAC Infrared Science Archive\footnote{http://irsa.ipac.caltech.edu/applications/DUST/}. The individual bandpass corrections were approximated from the corresponding SDSS filter corrections and thus have a minor systematic uncertainty not included in Table \ref{tab:Photom}.

\begin{table}
\begin{tabular}{@{}c c c c c c c}
\hline
MJD & $\Delta$T & Instrument & Filter & Magnitude & Seeing \\ [0.5ex]
 & (d) & & & & ($\prime\prime$) \\ [0.5ex]
\hline 
55745.1 & 17.1 & GMOS-S & $i^{\prime}$ & 22.10$\pm$0.10 & 1.4 \\
55749.0 & 21.0 & GMOS-S & $i^{\prime}$ & 21.96$\pm$0.10 & 1.4 \\
56274.3 & 546.3 & GMOS-S & $r^{\prime}$ & 23.74$\pm$0.17 & 0.7 \\
56274.3 & 546.3 & GMOS-S & $i^{\prime}$ & 22.76$\pm$0.12 & 0.7 \\
56277.3 & 549.3 & GMOS-S & $r^{\prime}$ & 23.60$\pm$0.26 & 0.8 \\
56284.2 & 556.2 & GMOS-S & $r^{\prime}$ & $>$22.84 & 0.9 \\
56536.0 & 808.0 & FORS2 & $z^{\prime}$ & $>$22.10 & 1.6 \\
56538.0 & 810.0 & FORS2 & $I$ & 23.28$\pm$0.25 & 1.5 \\
56538.0 & 810.0 & FORS2 & $z^{\prime}$ & 23.29$\pm$0.29 & 1.4 \\
56660.3 & 932.3 & GMOS-S & $i^{\prime}$ & $>$22.37 & 1.4 \\
\hline
\end{tabular}
\caption{{\em Swift} J1112-8238 optical photometry. Limits are stated to 3$\sigma$. Photometry is presented without host subtraction, although it is likely that the late epochs represent the host; that is, not significantly contaminated by transient light. Note the $i^{\prime}$ GMOS-S magnitudes were calculated using relative photometry from the VLT $I$-band image and so have a minor systematic uncertainty not included here. All observation times are measured from the beginning of the first day of the 4 day {\em Swift} trigger observation (2011 June 16 {\sc ut} 00:01). The seeing of each observation is included as it affects the contamination from the nearby bright star}
\label{tab:Photom}
\medskip

\end{table}

\subsection{Spectroscopy}\label{sec:spect}

Optical longslit spectroscopy of {\em Swift} J1112.2-8238 was obtained on GMOS-S on 2012 December 16 and 23 using the R400\_G5325 grating and independently on FORS2 using the 300I+11 grism on 2013 September 5. The GMOS-S spectra had a combined integration time of 2400 seconds (4 $\times$ 600) with spectral resolution of \mytilde7\AA\, and a spectral range of 3870 -- 8170\AA. The FORS2 spectrum also had an integration time of 2400 seconds (4 $\times$ 600) with spectral resolution of \mytilde12\AA\, and a spectral range of 5100--11000\AA.  The standard recommended Gemini {\sc iraf} and FORS2 {\sc esorex} data reduction was carried out on the appropriate spectra. 

\begin{figure*}
\includegraphics[width=16.8cm]{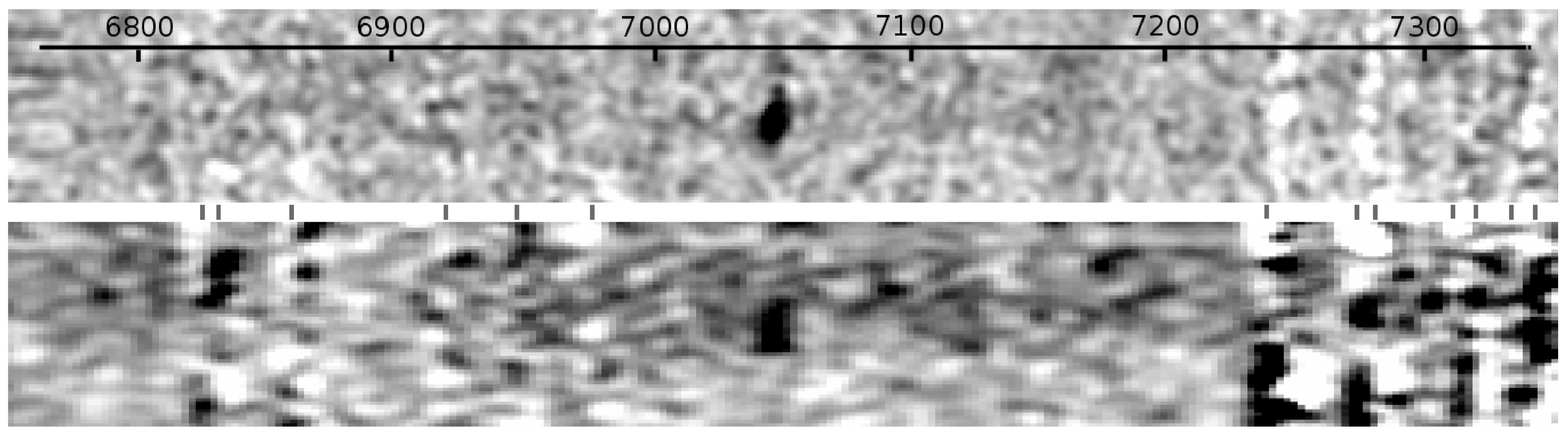}
 \caption{The (Top) GMOS and (Bottom) FORS2 spectra showing the clear emission line at a wavelength of $\mytilde7045$\AA, interpreted as the O{\sc ii} emission line, though the low resolution and line signal to noise preclude the possibility of resolving its doublet nature. The position of the removed sky lines are indicated as dashes between the two spectra. The scale is in units of Angstroms.}
\label{fig:Spectrum}
\end{figure*}

In all spectra, a single, weak emission feature was observed at $\mytilde 7045$\AA$\,$ (Figure \ref{fig:Spectrum}), with a significance of 
$\mytilde 10 \sigma$ in the GMOS spectrum. No continuum flux, or additional emission lines were seen. 
The line does not lie at the position of any common zero redshift features. It is offset by $\mytilde 600$ km s$^{-1}$ from the He 7060\AA\ line that is sometimes seen in accreting binaries \citep[][]{Marsh1991}. However, in these binaries the line is broad, and many other emission features are seen. In addition, the existence of an underlying extended source, interpreted as the host of the transient, greatly reduces the probability of a Galactic origin as nebulae are the only Galactic source likely to be resolvable, and these typically show multiple emission lines. This indicates that {\em Swift} J1112-8238 is not a Galactic source.

The non-detection of other lines proximate in wavelength disfavours the identification of this line as either [O{\sc iii}]($\lambda$4959,5007\AA) or H$\beta$ at $z \mytilde 0.4$, since in either case we would expect to observe the other lines. If the line were H$\alpha$ at $z=0.07$,
we may expect to observe either [N{\sc ii}$] \lambda 6584$, or 
H$\beta$ and [O{\sc iii}], since all lie within the spectral window covered by our GMOS observations. The expected H$\beta$ flux can be calculated directly
(under the assumption the observed line is H$\alpha$, and that the host galaxy extinction is minimal, as implied by the X-ray absorption). However, the combination of grating efficiency and Galactic reddening mean that we do not expect to observe H$\beta$ in our observations at $>1.5\sigma$. The [O{\sc iii}]
lines can frequently be substantially brighter than H$\beta$, and for a galaxy of metallicity $12+\log(\mathrm{O/H}) \mytilde 7.8$, consistent with the inferred absolute 
magnitude at $z=0.07$ \citep[rest frame $M_r \mytilde -14.8$,][]{Sweet2014}, we would expect [O{\sc iii}] ($\lambda$ 5007\AA) to be a factor $\mytilde 6$ brighter than H$\beta$. Accounting for foreground extinction and grating efficiency as 
before, we estimate we would expect to observe it at $\mytilde 8\sigma$, whereas no line is present at this location. Any emission at the location of  [N{\sc ii}] $\lambda 6584$ would be well below the detection limit given this assumed metallicity. We also note that at 
$z=0.07$ the absolute magnitude of the galaxy of $M_i > -15$ would be unusually faint. Given these combined constraints we disfavour 
the origin of the line as H$\alpha$. 

Hence we identify the line as O{\sc ii} ($\lambda$ 3727\AA) at a redshift $z=0.8901\pm0.0001$. In this case, the redward emission lines are beyond the range of our GMOS spectroscopy, and lie in bright sky lines in our FORS observations, precluding their detection. The low resolution of the spectra means that we are unable to resolve the doublet in this case. This interpretation is supported by the observed galaxy colours. After correction for foreground extinction they are relatively red in $r-i\, \mytilde\,0.9 \pm 0.2$, and bluer 
$i-z\, \mytilde\, -0.5\pm0.3$ (based on the $\mytilde550$ day Gemini $i^{\prime}$ band photometry). Although the errors are large, this is consistent with the presence of a Balmer break between the $r-$ and $i-$bands, as might be expected for $z=0.89$.

\section{Discussion}\label{sec:discussion}

Our spectroscopic observations have confirmed the existence of a single emission line that is inconsistent with any zero redshift lines and consistent with our adopted redshift, $z=0.8901\pm0.0001$. In this section we provide a short summary of the inferred rest-frame transient and host properties and then compare them to the properties of possible progenitors.

\subsection{Physical properties}

At the time of the first X-ray observations (10 days post trigger), the isotropic X-ray luminosity of the source was $\mytilde 6\times 10^{46}$erg\,s$^{-1}$. The source showed an approximate power law decay with time of t$^{-1.1}$. However this had considerable short-timescale variability superimposed upon it, with factor of 2 differences in flux on timescales of a few thousand seconds ($>10^6$s after the initial outburst). The X-ray spectrum was well fit by a power law spectrum with $\Gamma = 1.33$ and there was no evidence of spectral evolution. While multi-component fits to the time series data involving a broken power law or a flare produce statistically better fits, this may simply be due to the intrinsic short-term variability of the source and the sparse sampling of the X-ray lightcurve, precluding the inference of more detailed information about the source.

By assuming the late-time optical epochs represent host level flux that is uncontaminated by transient light, we can can
subtract this from our earlier observations. This was done through the use of the image subtraction software {\sc isis} \citep[][]{Alard1998}. We aligned, convolved and subtracted the late time ($\mytilde 1.5$ years) image from the early time (17 and 21 days) images in $i^{\prime}$. The subtractions left a clear, point source residual in each image with an inferred position that lay at 0.11$^{\prime\prime}\pm0.12^{\prime\prime}$ and 0.22$^{\prime\prime}\pm0.11^{\prime\prime}$ (1$\sigma$) from the centroid of the host galaxy, determined using a  S\'{e}rsic profile fit to the late-time image using the {\sc galfit} software package, as shown in Figure \ref{fig:centroid}. The error on the host centroid position is determined under the assumption of a Gaussian profile with FWHM equal to the half light radius from the {\sc galfit} S\'{e}rsic fit. The apparent asymmetry of the host means that this represents a lower limit on the true error in the centroid position. At the inferred redshift, our tightest constraint places the transient $0.85\pm0.93$\,kpc from the centre of its host, for which the half light radius is $\mytilde6$kpc.

The host subtraction also allows us to isolate the optical transient light, determining an absolute magnitude of M$_{{\mathrm{i}^{\prime}}}=-21.4$, equating to a luminosity of $\mytilde 1 \times 10^{43}$erg\,s$^{-1}$. The underlying host has a comparable absolute magnitude, M$_{{\mathrm{i}^{\prime}}}=-21.7$ (rest frame $M_g$ at $z=0.89$).
Based on the luminosity function of galaxies from \citet{Gabasch2006} this places it near $L^*$ at $z=1$ (at a redshift of 0.89, the $i^{\prime}$-band equates roughly to rest frame $g^{\prime}$-band for which, in the redshift range $0.85-1.31$, the $L^*$ magnitude is -21.7).

\subsection{Comparison to other sources}

\subsubsection{GRBs}
{\em Swift} detected gamma-ray bursts (GRBs), are typically detected on timescales much shorter than those for {\em Swift} J1112-8238. The majority arise from standard rate triggers, although a significant minority are longer-lived and trigger the detector via image triggers, sometimes on timescales of $>1000$ s. However, even the ultra-long GRBs \citep{Levan2014} that have durations of $\sim 10^4$ s are much
shorter than {\em Swift} J1112-8238, whose several day long $\gamma-$ray emission would imply a duration (if defined as $T_{90}$ as for GRBs)
of closer to $10^6$s. Hence on the basis of the $\gamma$-ray properties alone, {\em Swift} J1112-8238 is a much closer analog with
{\em Swift}J1644+57 and {\em Swift} J2058+0516 than with any identified population of GRBs. 

The X-ray properties are also apparently distinct, since the inferred isotropic X-ray luminosity lies an order of magnitude above GRBs at a similar epoch \cite[see e.g. ][]{Nousek2006,Levan2014}, and GRB afterglows at such late times seldom show such pronounced variability (likely due to the lack of engine activity). Despite the longevity of the gamma-ray emission in ULGRBs, their late time afterglows are generally consistent with, if not slightly fainter than \citep{Campana2011,Thone2011,Evans2014}, those of normal lGRBs, and so the X-ray properties also would suggest a physically distinct system. 

The optical properties of {\em Swift} J1112-8238 are rather less conclusive. The optical transient luminosity is comparable with the brightest end
of the GRB afterglow distribution \citep[e.g.][]{Kann2011}, although given the X-ray brightness the inferred X-ray to optical spectral slope is very flat
($\beta_{OX} \sim 0.14$). If the emission mechanisms were similar to GRBs this would identify the counterpart of {\em Swift} J1112-8238 as a dark burst, 
and would imply significant extinction \citep[][]{Fynbo2009,Perley2013}, the correction for which would make the afterglow the brightest seen. Alternatively, one may ascribe rather different emission mechanisms to the counterpart to {\em Swift} J1112-8238, in which case little extinction may be needed. It is 
interesting to note in this regard that {\em Swift} J2058+0516 also has a very flat $\beta_{OX} \sim 0.11$, despite a strong UV-SED that implied little extinction \citep{Cenko2012,Pasham2015}. 

Finally, one can also contrast the locations of {\em Swift} J1112-8238 with those of GRBs. Long GRBs tend to trace the brightest regions of their host \citep[][]{Fruchter2006} which, given the low spatial resolution of the optical images and small angular size of the galaxy, might show itself as a coincidence of the transient position and the host centroid. Indeed, in the study of \cite{Fruchter2006} approximately 1/6 of the bursts were consistent with the brightest pixels in their host galaxies at {\em HST} resolution. Since we currently lack such high resolution images the strength of the association of 
{\em Swift} J1112-8238 with its host nucleus is rather weaker than in the cases of {\em Swift} J1644+57 and {\em Swift} J2058+0516. However, in the majority of other regards its properties find a much better match with these events than either with normal long-GRBs, of the ultra-long GRB population.

\subsubsection{AGN}

The apparent coincidence of the transient position and the host centroid makes an association with the central supermassive black hole of the galaxy plausible, and therefore possibly with ongoing AGN activity. No catalogued source is consistent with the position of {\em Swift} J1112-8238 in either the SUMSS 843GHz survey \citep[60\% complete down to 6mJy, 100\% to 8mJy,][]{Bock1999,Mauch2003} or the AT20G 20GHz survey \citep[91\% complete to 100mJy,][]{Murphy2010}. This places limits on the pre-flare underlying radio emission of the host to the $10^{25}-10^{26}$\,W\,Hz$^{-1}$ level, which is only capable of ruling out the most luminous BL Lac type objects \citep[][]{March2013}. However, while the X-ray luminosity of the brightest blazar flares can reach the levels observed in {\em Swift} J1112-8238, this is generally accompanied by optical emission many magnitudes brighter than presented here, as seen in Figure \ref{fig:X-ray-Opt}. In addition, our late-time X-ray limit places constraints on any underlying activity to a limit of L$_X < 10^{44}$\,erg\,s$^{-1}$, fainter than the majority of quasars.  For these reasons, we disfavour the identification of this flare as a blazar flare.

\subsubsection{Relativistic Tidal Disruption Flares}

The association of the optical flare with the inferred location of the SMBH may indicate the discovery of a new tidal disruption flare.
In order to determine if this is plausible, we estimate the mass of the black hole expected to occur within a galaxy of this size. \citet[][]{Kauffmann2003} measure mass to light ratios of galaxies for a given redshift and rest-frame $g$-$r$ colour. At a redshift of 0.89, this equates roughly to a $i$-$z$ band colour in the observer frame. Based on an $i$-$z$ colour of $-0.5$ the mass to light ratio is $\mytilde$0.1, and, coupled with the rest-frame $g$-band (observer frame $i$-band) absolute magnitude of -21.7, implies a relatively high galaxy mass of $2\times10^{9}$M$_{\odot}$. The stellar mass to black hole mass scaling relation of \citet[][]{Bennert2011}, produces a lower estimate for the SMBH mass of $\mytilde 5 \times 10^{6}$M$_{\odot}$ (although there is considerable scatter in this relation and it is unclear whether the relation is applicable to such low masses) which is approximately consistent with the result obtained using the method from \citet[][]{Haring2004} of $\mytilde2\times10^6$ (by assuming that the stellar mass estimate represents an upper limit on the bulge mass of the host). Both of these latter estimates are well within the $10^8$M$_{\odot}$ limit for a Sun-like star to be disrupted by a SMBH and produce a visible TDF, making a TDF origin plausible.

\begin{figure*}
 \includegraphics[width=16.8cm]{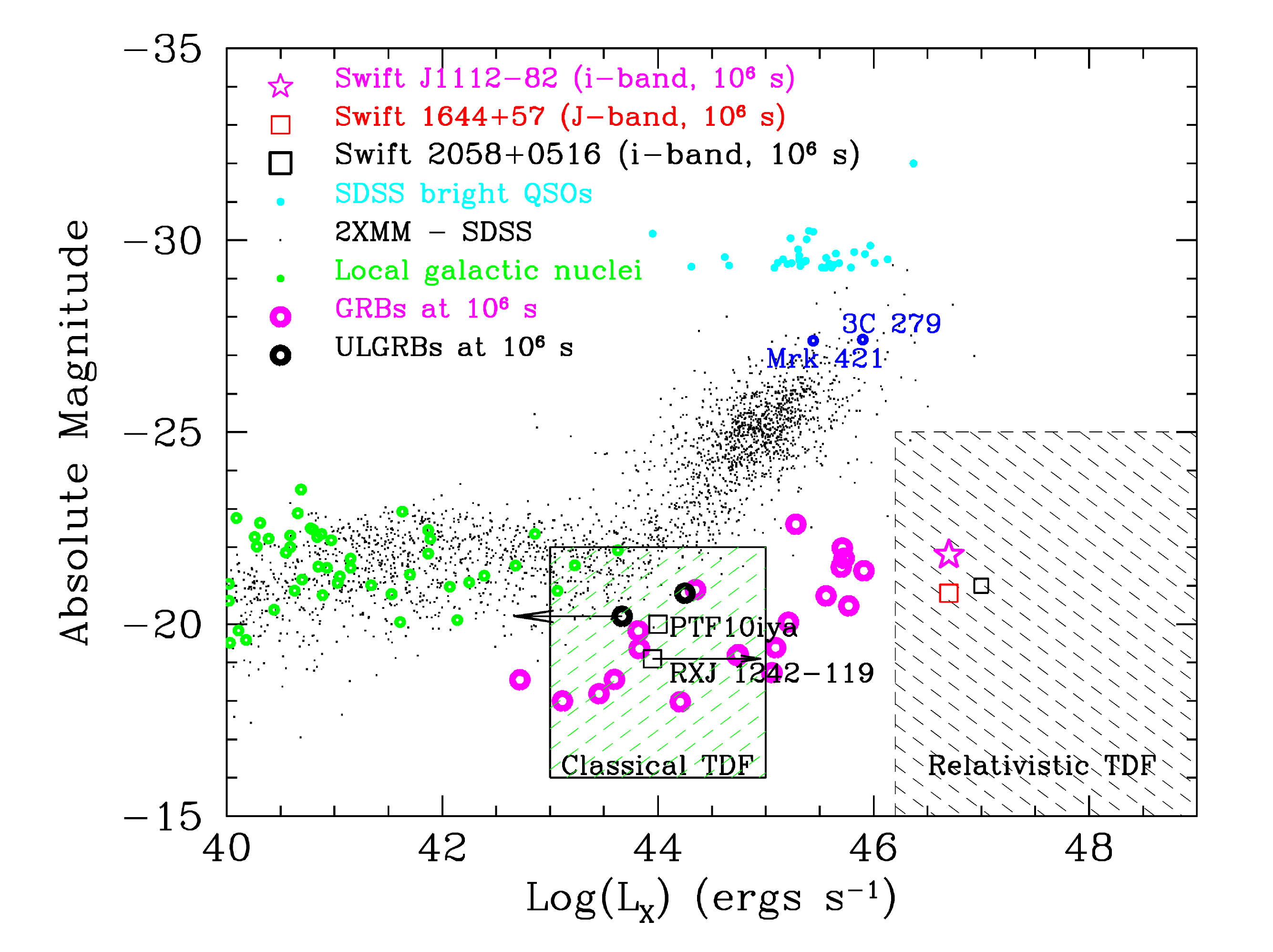}
 \caption{The X-ray luminosity and optical absolute magnitude plotted for a number of extragalactic transients, including AGN flares and GRBs at late times. {\em Swift} J1112-8238 is more X-ray luminous at $\mytilde10^6$ seconds than GRBs at similar epochs and, while the brightest X-ray blazar flares can match it, {\em Swift} J1112-8238 is very optically underluminous in comparison. It instead occupies a region of the parameter space devoid of other sources except {\em Swift} J1644+57 and {\em Swift} J2058+05. At early times the luminosity of these flares exceeds $10^{48}$ erg s$^{-1}$, and is in excess of $10^{46}$ erg s$^{-1}$ at $10^6$ s. This is more luminous than ``classical" TDFs, which exhibit markedly lower X-ray luminosity ($10^{44}$ erg s$^{-1}$), although none of these have been observed close to peak.  Adapted from \citet[][]{Levan2011}. \label{fig:X-ray-Opt}}
\end{figure*}

\begin{figure}
 \includegraphics[width=8.4cm]{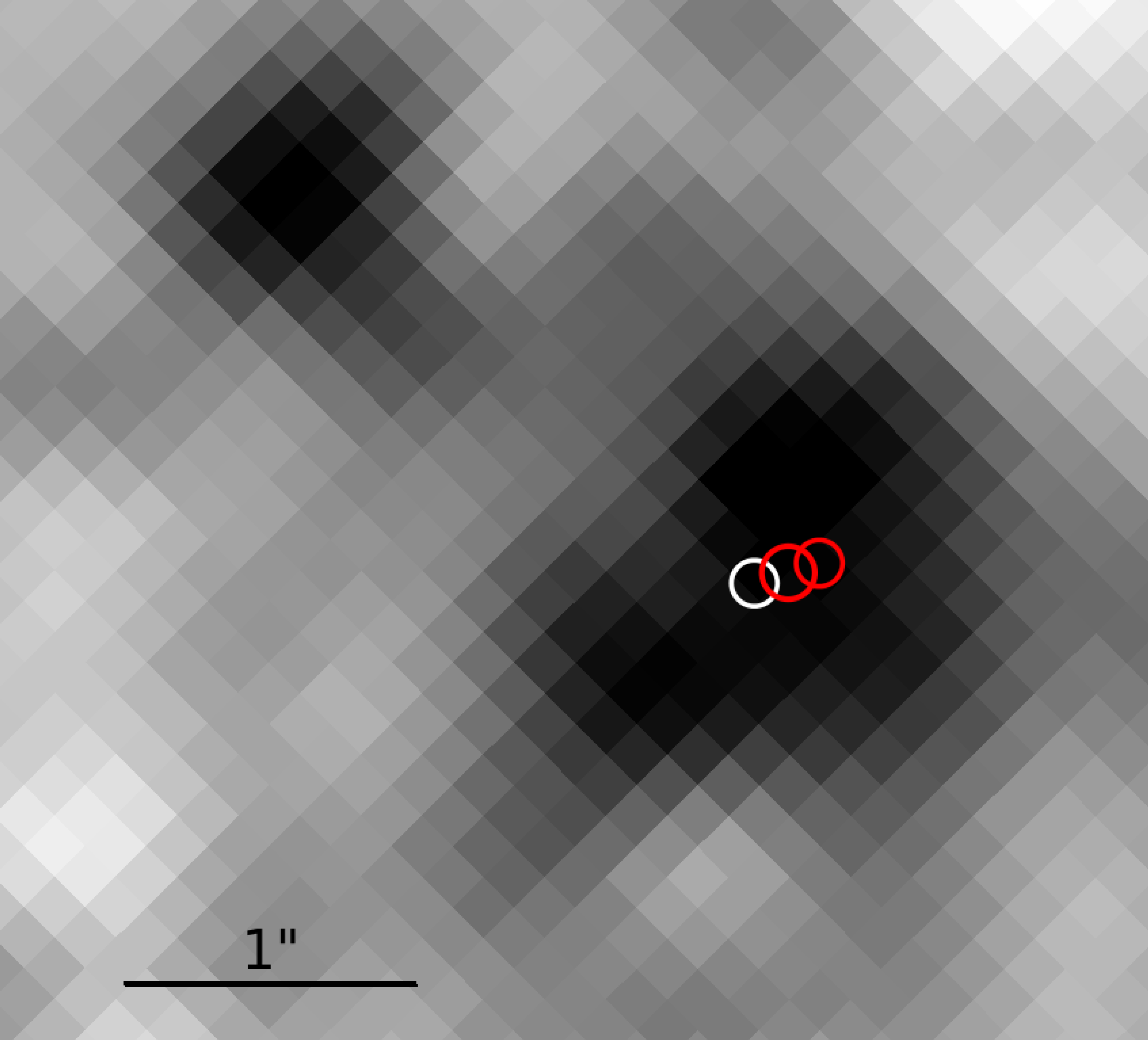}
 \caption{A zoomed in region ($\mytilde 5^{\prime\prime}$, North up, East left) around {\em Swift} J1112-8238, $\mytilde$ 550 days after initial detection, smoothed via a 3 pixel Gaussian convolution for clarity. The positions of the optical transient centroid, as measured 17 (left) and 21 (right) days after trigger with the {\sc iraf} command {\sc imexam}, are displayed as red 1-$\sigma$ error circles. Similarly, the host optical centroid (white) as measured via a S\'{e}rsic profile fit with {\sc galfit} is plotted as a white 1-$\sigma$ error circle. This error is a lower limit based on the assumption of a Gaussian profile with FWHM equal to twice the half-light radius of the S\'{e}rsic fit. The transient positions are thus coincident with the central position to 1$\sigma$ and 2$\sigma$ respectively. The consistency of these positions makes an association of the event with the SMBH in the galaxy plausible. It should be noted, though, that due to the low surface brightness and possible complex morphology of the host galaxy the host centroid is subject to substantial systematic uncertainty. \label{fig:centroid}}
\end{figure}

From Figure \ref{fig:X-ray-Opt}, we note that the optical absolute magnitude and X-ray luminosity of {\em Swift} J1112-8238 places it an region of phase space that is devoid of any sources with the exception of the aforementioned relativistic TDF candidates {\em Swift} J1644+57 and {\em Swift} J2058+05. These candidates also match well with this new flare in their late-time X-ray lightcurves as shown in Figure \ref{fig:lightcurve}, particularly in the case of {\em Swift} J1644+57. The overall power law decay observed over the 30 days of {\em Swift}-XRT follow-up of {\em Swift} J1112-82 is somewhat shallower than that of the other candidates with an index of $\mytilde -1.1$. {\em Swift} J2058+05 had a much steeper decay at a similar epoch with an index of $\mytilde -2.2$, while {\em Swift} J1644+57 had a late-time decay remarkably close to the t$^{-5/3}$ relation suggested to be a feature of TDF lightcurves \citep[][]{Rees1988,Phinney1989}.

However, this decay index is somewhat sensitive to the choice of $T_0$, which in this case is poorly defined, due in part to the unusual trigger method. Further, while often $T_0$ is taken to be the time at which the flare becomes observable, the true $T_0$ occurs some time earlier at the point of return of the most bound material which may precede emission by several days. In order to be consistent with a $t^{-5/3}$ decay, the ``true'' $T_0$ would have to have been $12^{+6}_{-4}$ before the start of the {\em Swift} detection image. This may not be unreasonable, since 
{\em Swift} J1644+57 was active at least 4 days prior to its first GRB trigger, and had a $3\sigma$ detection on a single day, 14 days earlier \citep{Krimm2011GCN}.  Constraints on $T_0$ have been attempted in detailed models of previous flares \citep[e.g. ][]{Guillochon2014}, however the lack of comprehensive follow-up precludes that possibility in this case. Perhaps even more importantly, the short duration over which observations were made also makes it difficult to determine the behaviour of the lightcurve within the context of the longer term emission. Indeed, {\em Swift} J1644+57's lightcurve was relatively flat at a similar epoch. Calculations considering more detailed transport of material through the disc point to a more complex picture, in which the t$^{-5/3}$ decline is only present in certain bands and over a rather restricted range of time \citep[][]{Lodato2011}, while even more recent calculations suggest that the $t^{-2.2}$ decay seen in {\em Swift} J2058+0516 should be present
in half of disruptions \citep{guillochon2013}. These predictions show that the X-ray flux can plateau over a period of tens of days after the initial disruption meaning the shallow decay of {\em Swift} J1112-8238 cannot place strong constraints on its nature. However it should be noted that these simulations concern the disk emission, whereas, in relativistic TDFs, the X-ray emission is thought to be dominated by the jet. It is unclear if the assumption of a direct correlation between the jet and disc emission is reasonable.

 Spectrally, the low number of counts recorded in {\em Swift} J1112.2-8238 restricts the information that can be extracted. However, the 
spectrum is well fit with a single, absorbed power-law with a relatively hard spectral index $\Gamma = 1.33 \pm 0.08$, (w-stat/dof = 574/586). This
is somewhat harder than the late time power-law index in {\em Swift} J1644+57 ($\Gamma \mytilde 2$) or in {\em Swift} J2058+0516 ($\Gamma \sim 1.6$).
 One area in which previous rTDFs differ is in 
the apparent correlation between hardness and flux. {\em Swift} J1644+57 exhibits spectral softening as it fades \citep{Levan2011}, while {\em Swift} J2058+0516 appears
to harden \citep{Cenko2012}. For {\em Swift} J1112.2-8238 it is not possible to discern any variation in spectrum with flux level. 

The rapid variability observed in the X-ray emission can place constraints on the nature of the emission region. While the variability is not as dramatic as that observed in {\em Swift} J1644+57, where factor of 100 changes in flux were observed on timescales of $\mytilde 100$ seconds, there is still evidence for factor of 2 variability on timescales of a few thousand seconds. Unfortunately the brightness of the source precludes timing at much higher resolution, and so light-travel time arguments would only place weak constraints on the size of the emitting region ($<10^{15}$cm, or 100 $R_S$ for a $\sim 10^7$ M$_{\odot}$ black hole). More compellingly, the gamma ray emission at the time of the first XRT observations  is close to the Eddington luminosity of a 10$^{9}$\,M$_{\odot}$ black hole, and an extrapolation to early times suggests it was brighter still. The expected black hole mass is a factor of several hundred small than this, it is unlikely that a black hole could accrete at such high super-Eddington rates, so while the constraints on beaming are weaker than for {\em Swift} J1644+57 we still believe this is the most likely explanation for {\em Swift} J1112-8238.

The lack of more comprehensive optical follow-up precludes the building of an optical SED which would help distinguish between the thermal SEDs of previous TDFs, (e.g. ASASSN-14ae, \citet{Holoien2014}; PS1-10jh, \citet{Gezari2012}), for which the peak absolute magnitudes are loosely consistent, and the differing, non-thermal emission mechanisms suggested in \citet[][]{Burrows2011} and \citet[][]{Bloom2011} for rTDF candidates. One of these models involves a blazar-analogue combination of inverse Compton emission at high frequencies (X-ray/$\gamma$) with a second peak at low frequencies (Optical etc.) from synchotron emission. Alternatively the emission in different wavebands may come from spatially separate emission regions. In the lightcurves of {\em Swift} J1644+57, limits on optical/radio short-term variability set the emission apart from the rapidly varying high energy emission. Under the assumption of a spherical emitting region with a blackbody temperature of $10^{5}$\,K ($10^{4}$\,K), the radius of the region emitting optical light in {\em Swift} J1112.2-8238 would be about $\mytilde3\times10^{15}$\,cm ($\mytilde1\times10^{16}$\,cm). This is approximately consistent with 3 (10) times the tidal radius of a of Sun-like star around a $10^{7}$\,M${_\odot}$ black hole. This result is  similar to those obtained from analysis of optical TDFs, perhaps unsurprisingly as the optical luminosity of {\em Swift} J1112-82 is only a factor of a few higher than seen in some other TDFs \citep[e.g.][]{vanVelzen2011,Gezari2012}. This may suggest a common mechanism for the optical emission from both relativistic and thermal TDFs.

It is also interesting to note {\em Swift} J1112-8238 shows a sharp decline in its X-ray flux at $\mytilde40$ days post trigger. This may be indicative of dipping as seen in {\em Swift} J1644+57 \citep{Levan2011,Bloom2011,saxton2012} at similar times, or perhaps of a longer term cessation of activity as identified in {\em Swift} J1644+57 at much later epochs of $\mytilde 1.5$ years 
\citep{2012ATel.4398....1S,levantanvir2012,Berger2012} and similarly in {\em Swift} J2058+05 \citep[][]{Pasham2015}. In any case, the final epoch of observations results in a limit which is significantly below the extrapolation of the early emission, requiring either a steepening of the decay or a rapid drop. This suggests broad similarities between the different events, although the sparse sampling of {\em Swift} J1112-8238, makes it difficult to rule out alternate interpretations.

With the previous rTDF candidates, a variable radio source with a measured Lorentz factor of $\mytilde2$ or higher was detected \citep[][]{Zauderer2011,Cenko2012}. In addition the inferred formation epoch from {\em Swift} J1644+57 implied a recently formed source, consistent with the start of the higher energy emission. This helped lead to the suggestion of a newly formed relativistic jet that accompanied the TDF. The lack of post-flare radio data precludes a similar analysis in the case of {\em Swift} J1112-8238. However, a {\em Swift} J1644+57-like radio lightcurve would be observable even several years after the flare and thus radio constraints may yet be obtainable.

\section{Implications }\label{sec:implications}

If {\em Swift} J1112-8238 is indeed a member of the same class of object as {\em Swift} J1644+57 and {\em Swift} J2058+0516 then it brings the total number of such events, as selected
by the high energy emission, to three. Radio observations of thermal (non-relativistic) TDF candidates \citep[e.g.][]{Bower2011,Bower2013,vanVelzen2013} have been used to attempt to determine the number of ``off-axis'' members and in the case of \citet[][]{Bower2013}, a few candidates may have been discovered. However, it is clear that the detected population is small. Nonetheless it is striking that these three outbursts were all discovered by {\em Swift} in the space of a 3 month window in 2011. At first sight
it may be argued that the proximity, and consequent brightness, of {\em Swift} J1644+57 may have motivated the searches that led to the discoveries of the additional candidates. However, the lack of any further examples in the subsequent four years suggests that this is more likely a statistical fluke. It is possible to quantify this via an archival search of {\em Swift} GRBs and the BAT transient monitor \citep{Krimm2013}. Within the 6.5 years of data reported in \citet{Krimm2013} there are two events marked as TDFs (the previously identified bursts), while only a further three are marked as ``unknown". Two of these ({\em Swift} J1713.4-4219, IGR J17361-4441) lie close to the Galactic plane, and are most likely Galactic sources. This leaves only the source under discussion, {\em Swift} J1112-8238, as a candidate relativistic TDF. It is plausible, though, that some other sources within the catalogue have been misidentified. In particular, {\em Swift} J1644+57 was initially identified as a Galactic Fast X-ray Transient \citep{Kennea2011}. However, the population detected by the BAT transient monitor is necessarily small. 

In total, therefore, it appears that at most a handful of such events have been recorded over {\em Swift}'s $\mytilde9$ year lifetime. Similar to \citet[][]{Cenko2012}, we can determine an implied analogous (i.e. similar {\em isotropic} luminosity) relativistic TDF rate based on the 3 events observed in $\mytilde$10 years, using the volume bounded by the distance to {\em Swift} J2058+05, as the most distant yet observed at $z = 1.1853$ giving a comoving volume of 215 Gpc$^3$, and assuming the local number density of $10^6$--$10^8 M_{\odot}$ SMBHs to be $10^{-2}$ Mpc$^{-3}$ \citep[][]{Tundo2007}. The resulting rate is found to be $\mytilde 3\times10^{-10}$ per galaxy per year, in stark contrast to the $\mytilde10^{-5}$ inferred from thermal TDF detections \citep[e.g.][]{Donley2002,vanVelzen2014}. 
Even if a significant fraction of the ULGRB population were related to similar phenomena this would be unlikely to constitute the majority of the factor of $3\times10^4$ required. To resolve this discrepancy likely requires a combination of tightly beamed high energy emission, such as that seen in GRBs, and that not all TDFs produce relativistic jets. 

Recent late time radio surveys of thermal TDFs by \citet{Bower2013} suggest that $\mytilde10\%$ of TDFs may have an associated relativistic jet. Given this, the required beaming angle for rTDF high energy emission would be of order $1^{\circ}$. At first sight this is not unreasonable, given that, for example, {\em Swift} J1644+57 produced an isotropic X-ray emission equivalent to the Eddington luminosity of a $10^{10}$M$_{\odot}$ black hole in a galaxy that is only expected to contain an SMBH of $\mytilde10^6$ solar masses \citep{Levan2011,Bloom2011}. In this case beaming (either relativistic and/or geometric) of a factor $10^{4}$, or highly super-Eddington accretion would seem to be necessary. 
However, radio observations of {\em Swift} J1644+57 point to a rather modest Lorentz factor, that would be unlikely to result in
such strong collimation ($\Gamma \mytilde 2$, \cite{Zauderer2011}), unless the radio and high-energy emission regions are spatially separate, each with their own Lorentz factors. It is also possible that the survey of \citet{Bower2013} could be impacted by small number statistics and potential contaminants. One of the two detections made, RXJ1420.4+5334 has an uncertain host identification due to the large error in the X-ray flare position. The second, IC3599, may be an AGN \citep[][]{Grupe1995}, and has recently exhibited repeated flares, either due to repeated partial disruptions of the same star on an $\mytilde10$ year orbit \citep[][]{Campana2015} or due to ongoing AGN activity \citep[][]{Grupe2015}. Because of this, the suggested 10$\%$ jetted TDF fraction may be overestimated, which would explain the lack of detections in any of the other studies \citep[e.g. ][]{Arcavi2014}, thus further contributing to the apparent deficit of detected rTDFs. Clearly further observations of larger samples of sources across the electromagnetic spectrum are needed to resolve this question.

\section{Summary}

We have presented multi wavelength observations of {\em Swift} J1112-8238, pinpointing it to close to the nucleus of an otherwise quiescent galaxy at $z=0.89$. The high
X-ray luminosity of the source coupled with its relative optical faintness occupy a region of parameter space which is populated only by the candidate relativistic TDFs. Hence
we suggest that {\em Swift} J1112-8238 is the third candidate member of this class of events as detected by their high-energy emission. The discovery of such a small number of events over the lifetime of {\em Swift} suggests that
they are extremely rare and that, unless the beaming angles are extremely small (evidence for which may come from the extreme observed luminosities coming from events in hosts with small expected SMBHs), then their 
true astrophysical rates are also only a small fraction (about 1 in ${10^5}$) of the likely non-relativistic TDF rate. The factors which govern the production of a relativistic outflow associated
with a given TDF remain unclear, but highlight the needed for rapid dedicated follow-up of the rare examples when they are found.

\section*{Acknowledgements}
We thank the referee for a careful and thoughtful report that has improved this paper. GCB thanks the Midlands Physics Alliance for a PhD studentship. AJL thanks STFC for support under grant ID ST/I001719/1,
and the Leverhulme Trust for support via
a Philip Leverhulme Prize. 
Based on observations obtained at the Gemini Observatory, which is operated by the 
Association of Universities for Research in Astronomy, Inc., under a cooperative agreement 
with the NSF on behalf of the Gemini partnership: the National Science Foundation 
(United States), the National Research Council (Canada), CONICYT (Chile), the Australian 
Research Council (Australia), Minist\'{e}rio da Ci\^{e}ncia, Tecnologia e Inova\c{c}\~{a}o 
(Brazil) and Ministerio de Ciencia, Tecnolog\'{i}a e Innovaci\'{o}n Productiva (Argentina).
Based on observations made with ESO Telescopes at the La Silla Paranal Observatory under programme ID 089.B-0860.
We have made use of the ROSAT Data Archive of the Max-Planck-Institut für extraterrestrische Physik (MPE) at Garching, Germany.

\label{lastpage}
\end{document}